# Lyapunov-based analysis of functional stability for edge computing systems

Oleksii S. Bychkov

Department of Software Systems and Technologies, Faculty of Information Technologies, Taras Shevchenko National University of Kyiv, Kyiv, Ukraine

Correspondence: oleksiibychkov@knu.ua

## Abstract

Functional stability has been proposed as a per-function reliability concept for edge computing, in which the unit of analysis is the individual service rather than the whole system and the binary working/failed evaluation is replaced by a continuous quality function with per-function thresholds. The definition is descriptive: a strong form requires $Q_i^d(t) \geq Q_i^{\min}$ at every time, a weak form requires the threshold to be restored within a finite recovery time. Verifying these properties for a concrete edge orchestrator, migration policy, or distributed inference service requires a constructive method. This article connects functional stability to the direct Lyapunov method. The strong form is characterized through positive invariance of the admissible region; the weak form is connected to uniform ultimate boundedness and input-to-state stability with analytical recovery-time bounds. Switched dynamics from service migration and node failover are treated via common Lyapunov functions. Finite-horizon variants address mission-bounded edge workloads.

## 1. Introduction

Edge computing systems run services on resource-constrained nodes located near data sources, in operating regimes where partial component failures, intermittent backhaul connectivity, thermal throttling, peer-node churn and bursty load are part of the normal operational envelope rather than rare events [1, 2, 3, 4]. Classical reliability analysis evaluates such systems on a binary working/failed scale and at the system level, which does not capture differentiated quality requirements of services co-located on the same edge node, and does not separate architectures that share the same nominal availability but distribute their operating margin between disturbance tolerance and recovery time in different ways.

A recent framework of *functional stability* for edge systems [5] addresses these limitations by shifting the unit of analysis from the system to the individual function and by replacing the binary evaluation with a continuous quality function $Q_i^d(t)$ and a per-function quality threshold $Q_i^{\min}$. The framework distinguishes a strong form (the condition $Q_i^d(t) \geq Q_i^{\min}$ for every disturbance $d \in D$, every function $f_i$ and every $t \geq t_0(d)$) from a weak form, which requires the threshold to be restored within a finite recovery time $T_{\text{rec}}^i(d)$ after disturbance onset. Per-function parameters (disturbance tolerance $R_d^{i,k}$, recovery time $T_{\text{rec}}^i$, degradation depth $\Delta Q_i$, admissible degradation $\delta_i$) and aggregate metrics (cumulative quality $Q_i$, functional availability $A_i$, integral stability $\Psi \in [0,1]$) connect to the formal definitions through a sequence of theorems.

The definitions and the metric system of [5] are *descriptive*: they specify what functional stability means for an edge system and how to score architectures against a declared disturbance-class scope, but they do not provide constructive methods to verify the property for a concrete orchestrator, container migration policy, or distributed inference service. Two natural questions remain: given a dynamic model of an edge system, how is functional stability verified? Given a candidate controller for resource allocation, model switching, or service relocation, how is it synthesized so that functional stability is guaranteed with quantitative recovery-time bounds?

The standard tool for rigorous stability analysis of nonlinear systems under disturbances is the direct Lyapunov method, developed in the classical monographs of Khalil [6] and Slotine [7] and extended to disturbance settings through robust stability, uniform ultimate boundedness (UUB) [6, 7] and input-to-state stability (ISS) [8, 9]. These tools were developed for the analysis of equilibrium points, whereas functional stability is concerned with the trajectory of quality indicators staying inside an admissible band. The adaptation is non-trivial: the admissible region $\Omega_\delta$ is a set rather than a point, the unit of analysis is per-function rather than system-level, and the disturbance enters explicitly through structured parametric classes that map to edge-relevant events such as link loss, CPU overload and node loss.

A further difficulty arises from structural changes induced by edge-specific events. Failure of an edge node in a cluster, migration of a service between nodes, switch from cloud to local inference, or eviction of a tenant under resource pressure all change the underlying dynamic model in a discontinuous way. Such systems are formally treated as *switched* dynamical systems [10, 11, 12], for which a single Lyapunov function may not exist; the theory of multiple Lyapunov functions and common Lyapunov functions provides the right setting.

This article establishes the formal connection between functional stability of edge computing systems and the direct Lyapunov method. The contributions are:

1. A change of variable from the per-function quality $Q_i^d(t)$ to the per-function *degradation* $z_i(t) = \max\{0, Q_i^{\text{nom}} - Q_i^d(t)\}$ that brings functional stability into the language of bounded-trajectory analysis on a compact admissible box $\Omega_\delta \subset \mathbb{R}_+^n$ (Section 3.1). The positive-part construction ensures that trajectories with quality above the nominal value (which are not violations of functional stability) are not penalized by quadratic Lyapunov candidates.

2. A criterion for the strong form through positive invariance of $\Omega_\delta$, with a Lyapunov-derivative condition on the boundary $\partial\Omega_\delta$ (Section 3.2, Theorems 1 and 2). A corollary identifies the strong form with a “margin” through a strict negative-definite condition on the derivative.

3. Connections of the weak form to uniform ultimate boundedness (Section 3.3, Theorem 3) and to input-to-state stability (Section 3.4, Theorem 4) with piecewise analytical recovery-time upper bounds whose non-trivial branch has the form $k^{-1}\ln(\cdot)$, where $k$ is the convergence rate of the Lyapunov function along trajectories.

4. Treatment of switched dynamics arising from service migration, node failover and adaptive computation through common Lyapunov functions (Section 3.5, Theorem 5).

5. A unified sufficient condition for functional stability of edge systems with switched dynamics and external disturbances (Section 3.6, Theorem 6).

6. Finite-horizon variants for SLA-bounded and mission-bounded edge workloads, with conditions on the observation horizon $T$ versus the recovery budget (Section 3.7, Definition 5 and Theorems 7–9).

7. A table relating the per-function parameters of [5] to the parameters of the Lyapunov construction (Section 3.8, Table 1).

The article is organized along the IMRaD structure recommended by the journal. Section 2 reviews related work on Lyapunov methods and on dependability in edge and fog computing. Section 3 develops the constructive framework. Section 4 instantiates it for edge computing and presents two analytical case studies. Section 5 discusses the results and limitations. Section 6 concludes.

# 2. Related work

## 2.1. Lyapunov methods for nonlinear systems under disturbances

The direct Lyapunov method for ordinary differential equations is treated systematically by Khalil [6] and Slotine [7]. For disturbed systems, two extensions are particularly relevant. *Uniform ultimate boundedness* (UUB) captures asymptotic confinement of trajectories to a bounded set; the condition $\dot{V} \le -kV + \rho$ along trajectories (with $k > 0$ and intensity $\rho \ge 0$) implies entry into a sublevel set determined by the ratio $\rho/k$. *Input-to-state stability* (ISS), introduced by Sontag [8] and Jiang and Wang [9], formalizes the dependence of the trajectory on the disturbance through a gain function $\gamma \in \mathcal{K}$ and a transient term $\beta \in \mathcal{KL}$. Both are surveyed in the textbook treatments of [6, 7].

Barrier Lyapunov functions [13, 14] address the problem of keeping trajectories inside a prescribed region by making the Lyapunov function tend to infinity at the boundary of that region. These are conceptually close to functional stability: the prescribed region is the admissible region $\Omega_\delta$, and the barrier acts as a soft enforcement of $Q_i^d(t) \geq Q_i^{\min}$.

### 2.2. Switched systems and common Lyapunov functions

Switched systems are treated by Branicky [10], Liberzon [11] and Hespanha and Morse [12]. A switched system has the form $\dot{x} = f_{\sigma(t)}(x)$, where $\sigma(t)$ takes values in a finite set of *modes* and the dynamics changes discontinuously when $\sigma$ jumps. The stability problem is non-trivial: even when each mode is asymptotically stable, the switched system may be unstable for some switching signals. A sufficient condition for stability under *arbitrary* switching is the existence of a common Lyapunov function (CLF), i.e., a single function $V$ that is a Lyapunov function for every mode. The multiple Lyapunov function (MLF) approach of Branicky [10] relaxes this requirement at the cost of introducing constraints relating mode-specific Lyapunov functions at switching instants.

For edge computing, the relevant switching events are service migration between nodes, failover between primary and standby instances, and mode changes within an adaptive controller (e.g., switching between full-precision and quantized inference). These are studied in the edge orchestration literature [15, 16, 17, 18] from an operational perspective; the corresponding stability analysis appears to be largely absent.

### 2.3. Functional stability in edge and fog computing

The functional stability framework of [5] is developed for edge computing and includes the per-function definitions, the parameter system, the integral metric $\Psi$ with explicit dependence on the disturbance-class scope $K_i$, and an instantiation for typical edge scenarios. Other reliability concepts for edge and fog environments are reviewed in [19, 20, 21]. Surveys of fault-tolerant control [22, 23] place Lyapunov methods at the centre of constructive controller synthesis under faults, with the focus on tracking error rather than per-function quality.

A gap remains between the descriptive functional stability framework of [5] and the constructive Lyapunov methods of [6, 7, 8, 9, 10, 11, 12]. The present article closes that gap.

## 3. Methods

### 3.1. Dynamic model of an edge system with quality functions

To apply Lyapunov methods, the per-function quality trajectories $Q_i^d(t)$ of [5] are linked to a dynamic state model. Let the state of the edge system be $x(t) \in \mathbb{R}^m$, and let its dynamics under disturbance $d(t)$ be

$$\dot{x}(t) = f(x(t), d(t)), \qquad d(t) \in D,$$

where $D = \bigcup_{k \in K} D_k$ is the structured set of disturbances with parametric classes $D_k = \{d_k(a): a \in [0, A_k^{\max}]\}$ introduced in [5]. The state vector $x$ collects all internal variables that drive the quality of the services co-located on the edge node: queue lengths, resource utilizations, cached weights, link buffer occupancies, and similar quantities. The per-function quality $Q_i$ is a measurable function of the state,

$$Q_i(t) = g_i(x(t)), \qquad g_i: \mathbb{R}^m \to [0, Q_i^{\max}].$$

For analysis purposes it is convenient to operate on *degradation* rather than on quality. Define the raw error

$$e_i(t) = Q_i^{\text{nom}} - Q_i(t) = Q_i^{\text{nom}} - g_i(x(t)),$$

and its *positive part*,

$$z_i(t) = \max\{0, e_i(t)\} = \max\{0, Q_i^{\text{nom}} - Q_i(t)\},$$

which represents the actual degradation from the nominal trajectory. The use of $z_i$ rather than $e_i$ matters: trajectories with $Q_i(t) > Q_i^{\text{nom}}$ ($e_i < 0$) correspond to quality above the nominal value and are not violations of functional stability; only the part $e_i > 0$ counts as degradation. Let $z(t) = (z_1(t), \dots, z_n(t)) \in \mathbb{R}_+^n$ be the degradation vector. The functional stability requirement $Q_i^d(t) \geq Q_i^{\min}$ rewrites as

$$z_i(t) \leq \delta_i, \qquad \delta_i = Q_i^{\text{nom}} - Q_i^{\min} > 0,$$

where $\delta_i$ is the admissible degradation introduced in [5]. The vector form

$$\Omega_\delta = \{z \in \mathbb{R}^n: 0 \leq z_i \leq \delta_i \text{ for all } i = 1, \dots, n\}$$

defines the *admissible region*: a compact axis-aligned box in $\mathbb{R}_+^n$ that contains exactly the degradation vectors compatible with functional stability. The strong form of functional stability is the requirement that the trajectory $z(t)$ remains inside $\Omega_\delta$ at every time; the weak form is the requirement that the trajectory returns into $\Omega_\delta$ within a finite time after every disturbance.

**Standing assumptions.** Throughout the article, $f$ is locally Lipschitz in $x$ uniformly in $d$, $g_i$ is continuously differentiable, and the disturbance set $D$ is bounded. These conditions hold for the queueing, fluid and resource-control models used in edge orchestration [15, 16, 17]. The zero-amplitude convention of [5] is preserved: $a = 0$ in any disturbance class corresponds to the absence of degradation, so the nominal trajectory $z \equiv 0$ is an equilibrium under no disturbance. The positive-part construction in (4) is non-smooth at $e_i = 0$; in what follows, statements about $\dot{V}$ along trajectories are understood in the Dini sense at $z_i = 0$, but for all trajectories of interest (those with positive degradation under a disturbance) the smooth restriction $z_i = e_i > 0$ applies, and standard smooth analysis is used.

### 3.2. Strong form via positive invariance

The strong form requires $z(t) \in \Omega_\delta$ for every $t \geq t_0(d)$. The natural mathematical tool is the theory of positively invariant sets [24].

**Definition 1 (Positive invariance).** *A set $\Omega \subset \mathbb{R}_+^n$ is positively invariant for the degradation dynamics $\dot{z} = h(z, d)$ if for every initial state $z(0) \in \Omega$ and every admissible disturbance $d(t) \in D$ the solution satisfies $z(t) \in \Omega$ for all $t \geq 0$.*

**Theorem 1 (Strong form ↔ positive invariance).** *The system (1)–(2) satisfies the strong form of functional stability with respect to $D$ if and only if the admissible region $\Omega_\delta$ is positively invariant for the degradation dynamics induced by (1)–(4), starting from any initial state $z(t_0) \in \Omega_\delta$.*

*Proof.* The strong form requires $Q_i^d(t) \geq Q_i^{\min}$ for every $t \geq t_0(d)$, which by (3)–(5) is equivalent to $z(t) \in \Omega_\delta$ for every $t \geq t_0(d)$. Positive invariance of $\Omega_\delta$ is therefore exactly the strong form when restricted to initial states inside $\Omega_\delta$. The converse is immediate. ▫

The admissible region (6) is a box, not a sublevel set of a smooth function. Two constructive sufficient conditions follow, corresponding to two ways of certifying invariance of a box.

**Theorem 2A (Component-wise barrier condition).** *The admissible region $\Omega_\delta$ defined in (6) is positively invariant for the degradation dynamics if for every $i = 1, \ldots, n$ and every $d \in D$,*

$$\dot{z}_i \leq 0 \quad \text{whenever } z_i = \delta_i \text{ and } z_j \in [0, \delta_j] \text{ for } j \neq i.$$

*That is, at every point of $\partial\Omega_\delta$ where the $i$-th face is active, the degradation $z_i$ does not increase further.*

*Proof.* The boundary $\partial\Omega_\delta$ is the union of the $n$ faces $\{z: z_i = \delta_i\}$. If condition (7) holds on every face, the inward velocity field on $\partial\Omega_\delta$ prevents trajectories from leaving $\Omega_\delta$ through any face. Suppose by contradiction that some trajectory leaves $\Omega_\delta$ at $t^* > t_0$; then by continuity there is a face index $i$ and a time $\tau \leq t^*$ with $z_i(\tau) = \delta_i$ and $\dot{z}_i(\tau) > 0$, contradicting (7). ▫

A second sufficient condition, useful when a quadratic Lyapunov function is constructed on the degradation variables, expresses invariance through a sublevel set of $V$ contained in $\Omega_\delta$.

**Theorem 2B (Lyapunov sublevel-set sufficient condition).** *Let $V: \mathbb{R}_+^n \to \mathbb{R}_+$ be continuously differentiable on $\operatorname{int}(\mathbb{R}_+^n)$ with $V(0) = 0$. Define the sublevel set $\Omega_V = \{z \in \mathbb{R}_+^n: V(z) \leq c\}$ with $c > 0$ chosen so that $\Omega_V \subseteq \Omega_\delta$. If along the trajectories of the degradation dynamics*

$$\dot{V}(z) = \nabla V(z) \cdot \dot{z} \leq 0 \quad \text{for every } z \in \partial\Omega_V \text{ and every } d \in D,$$

*then $\Omega_V$ is positively invariant. In particular, if the initial degradation satisfies $V(z(t_0)) \leq c$, the strong form holds.*

*Proof.* Standard sublevel-set invariance argument: outward crossing of $\partial\Omega_V$ at any $\tau > t_0$ would force $\dot{V}(z(\tau)) > 0$, contradicting (8). Containment $\Omega_V \subseteq \Omega_\delta$ then yields the strong form for trajectories starting in $\Omega_V$. ▫

**Quadratic Lyapunov candidate.** For the degradation variables $z = (z_1, \ldots, z_n) \in \mathbb{R}_+^n$, the standard quadratic candidate

$$V(z) = \sum_{i=1}^{n} \alpha_i \, z_i^2, \qquad \alpha_i > 0,$$

defines an ellipsoidal sublevel set. The largest ellipsoid contained in the box $\Omega_\delta$ has the level

$$c_\delta = \min_{i=1,\dots,n} \alpha_i \delta_i^2,$$

and is touched by $\Omega_\delta$ along the axis $z_i = \delta_i$ for the index attaining the minimum. The conservative invariance bound $\Omega_V = \{V \le c_\delta\} \subseteq \Omega_\delta$ is therefore tight along that axis but strictly inside $\Omega_\delta$ on the other faces. Theorem 2B applied with this $c_\delta$ gives a constructive sufficient condition for the strong form. Theorem 2A is tighter (it does not require $\Omega_V \subseteq \Omega_\delta$ but works directly with the box) at the cost of $n$ scalar inequalities rather than a single Lyapunov decrease.

**Corollary 1 (Strong form with margin).** *If the inequality (8) holds in the stronger form*

$$\dot{V}(z) \le -kV(z) \quad \text{for every } z \in \Omega_V \text{ and every } d \in D, \quad k > 0,$$

*then $\Omega_V$ is positively invariant; moreover trajectories starting in $\Omega_V$ converge exponentially to $z = 0$ with rate $k$. The strong form holds with a margin: $z_i(t) \le \delta_i$ at every time, and $z(t) \to 0$ as $t \to \infty$.*

The parameter $k$ in (11) plays a central role throughout the article: it is the convergence rate of the Lyapunov function and, as Theorem 3 below shows, sets the recovery-time scale of the weak form.

### 3.3. Weak form via practical stability (UUB)

The weak form of functional stability allows temporary violation of the threshold $Q_i^{\min}$ but requires return within a finite recovery time $T_{\text{rec}}^i(d)$. The Lyapunov-side counterpart is *practical stability*, also known as uniform ultimate boundedness [6]. To avoid clash with the admissible degradation $\delta_i$, the UUB intensity parameter is denoted $\rho$.

**Definition 2 (UUB).** *The degradation dynamics is uniformly ultimately bounded with parameters $k > 0$ and $\rho \ge 0$ if there exists a continuously differentiable Lyapunov function $V: \mathbb{R}_+^n \to \mathbb{R}_+$ such that*

$$\dot{V}(z) \le -kV(z) + \rho \quad \text{along trajectories,} \quad \text{for every } d \in D.$$

Solving the differential inequality (12) gives

$$V(z(t)) \le V(z(t_0))\, e^{-k(t-t_0)} + \frac{\rho}{k}\left(1 - e^{-k(t-t_0)}\right),$$

so $V(z(t)) \to \rho/k$ as $t \to \infty$. Trajectories enter and remain in the ultimate-bound sublevel set $\{z: V(z) \le \rho/k\}$.

**Theorem 3 (UUB ⇒ weak form with explicit $T_{\text{rec}}$).** *Suppose the degradation dynamics admits the quadratic Lyapunov candidate (9) and satisfies the UUB condition (12) with $k > 0$ and $\rho \ge 0$. Let $c_\delta = \min_i \alpha_i \delta_i^2$ be the largest level for which $\{V \le c_\delta\} \subseteq \Omega_\delta$ (cf. (10)). If*

$$c_\delta > \frac{\rho}{k},$$

*then the system satisfies the weak form of functional stability, with the recovery-time upper bound $T^i_{\mathrm{rec}}(d) \le B_{\mathrm{rec}}(t_0)$, where*

$$B_{\mathrm{rec}}(t_0) = \begin{cases} 0, & \text{if } V(z(t_0)) \le c_\delta, \\ \dfrac{1}{k}\ln\dfrac{V(z(t_0)) - \rho/k}{c_\delta - \rho/k}, & \text{if } V(z(t_0)) > c_\delta. \end{cases}$$

*The value $B_{\mathrm{rec}}(t_0)$ is an upper bound on $T^i_{\mathrm{rec}}(d)$, not the exact recovery time: the realized trajectory may enter $\Omega_V$ earlier than the Lyapunov decay rate guarantees.*

*Proof.* Condition (14) ensures that the ultimate-bound sublevel set $\{V \le \rho/k\}$ is strictly contained in $\{V \le c_\delta\} \subseteq \Omega_\delta$. By (13), the trajectory enters $\{V \le c_\delta\}$ as soon as

$$V(z(t_0))\, e^{-k(t-t_0)} + \frac{\rho}{k}\left(1 - e^{-k(t-t_0)}\right) \le c_\delta,$$

which rearranges to $e^{-k(t-t_0)} \le (c_\delta - \rho/k)/(V(z(t_0)) - \rho/k)$ and yields the second branch of (15) as an upper bound on the entry time into $\Omega_V$. If $V(z(t_0)) \le c_\delta$, the trajectory is already inside the inscribed ellipsoid $\Omega_V \subseteq \Omega_\delta$ at $t_0$, the strong form holds from $t_0$ onwards under the UUB dynamics (since $\Omega_V$ is forward-invariant by Theorem 2B applied to the UUB-decrease condition on $\partial\Omega_V$), and the recovery time is bounded by 0. The piecewise form of $B_{\mathrm{rec}}(t_0)$ captures both cases without invoking the logarithm at non-positive arguments. After entry into $\Omega_V$, the trajectory stays in $\Omega_V \subseteq \Omega_\delta$ and the weak form holds for every $i$. ▫

**Remark 1 (Operational reading).** The parameter $\rho$ in (12) models the intensity of persistent disturbances on the edge node (background load, jitter, thermal noise, residual interference), while $k$ models the speed of the corrective control loop (orchestrator response, autoscaling latency, batch-renormalization frequency). The ratio $\rho/k$ is the *residual degradation budget* of the system: how far the steady-state degradation can deviate from the nominal trajectory. Condition (14) requires this budget to fit strictly inside the largest ellipsoid contained in $\Omega_\delta$.

**Remark 2 (Relation to Definition 4 of [5]).** Theorem 3 produces an upper bound on the per-function recovery time $T^i_{\mathrm{rec}}$ defined in [5]; the actual recovery time depends on the realized trajectory and may be smaller. Bound (15) is sufficient for the weak form and is conservative because it uses the level $c_\delta$ (the inscribed ellipsoid) rather than the box $\Omega_\delta$ itself.

### 3.4. Weak form via input-to-state stability (ISS)

The UUB characterization of Section 3.3 takes the bound $\rho$ on the disturbance effect as given. The ISS approach [8, 9] gives a more refined picture by making the dependence of the trajectory on the disturbance amplitude explicit.

**Definition 3 (ISS).** *The degradation dynamics is input-to-state stable if there exist* $\beta \in \mathcal{KL}$ *and* $\gamma \in \mathcal{K}$ *such that for every initial state* $z(0)$ *and every bounded disturbance* $d(\cdot)$*,*

$$|z(t)| \le \beta(|z(0)|, t) + \gamma\left(\sup_{\tau \in [0,t]} |d(\tau)|\right).$$

The function $\beta$ describes the decay of the initial-condition effect; $\gamma$, called the *gain*, describes the asymptotic effect of disturbance amplitude. ISS implies UUB but is strictly stronger: it gives quantitative bounds on both the transient and the steady-state response.

Let $\delta_{\min} = \min_i \delta_i$ denote the tightest admissible degradation across functions. For a parametric disturbance class $d_k(a) \in D_k$ with amplitude bound $A_k(a) = \sup_t |d_k(a)(t)|$ (assumed monotone non-decreasing in $a$, as in [5]), the ISS bound (16) gives the following.

**Theorem 4 (ISS ⇒ weak form with explicit $T_{\text{rec}}$ and $R_d^{i,k}$).** *Suppose the degradation dynamics is ISS with functions* $\beta$ *and* $\gamma$*. For a disturbance class* $d_k(a) \in D_k$ *with amplitude bound* $A_k(a)$*, if*

$$\gamma(A_k(a)) < \delta_{\min},$$

*then the weak form holds with recovery-time bound*

$$T_{\text{rec}}^i(d_k(a)) \le \inf\{t \ge 0\colon\ \beta(|z(t_0)|, t) \le \delta_{\min} - \gamma(A_k(a))\}.$$

*If, in addition, the disturbance is applied from the nominal degradation state* $z(t_0) = 0$*, the transient term* $\beta(|z(t_0)|, t) = \beta(0, t) = 0$*, and the strong form holds for the class* $D_k$ *at amplitude* $a$ *whenever* $\gamma(A_k(a)) \le \delta_i$*. Under monotonicity of* $A_k$ *in* $a$*, the per-class strong-form disturbance tolerance of [5] then satisfies the analytical lower bound*

$$R_d^{i,k} \ge A_k^{-1}(\gamma^{-1}(\delta_i)) \ge A_k^{-1}(\gamma^{-1}(\delta_{\min})).$$

*Proof.* By (16), for $t$ such that $\beta(|z(t_0)|, t) \le \delta_{\min} - \gamma(A_k(a))$, we have $|z(t)| \le \delta_{\min}$, hence $z_i(t) \le \delta_i$ for every $i$, so the strong form holds from that time on. Because $\beta(r, t)$ is non-increasing in $t$ for fixed $r$, the infimum in (18) is well-defined and finite under (17); the strict inequality in (17) ensures the right-hand side of (18) is strictly less than infinity. For (19), under $z(t_0) = 0$ the bound (16) reduces to $|z(t)| \le \gamma(A_k(a))$, so the strong form on class $D_k$ with amplitude $a$ holds whenever $\gamma(A_k(a)) \le \delta_i$, i.e., $a \le A_k^{-1}(\gamma^{-1}(\delta_i))$ under monotonicity. The supremum definition of $R_d^{i,k}$ in [5] then yields the lower bound. ▫

**Remark (nonzero initial state).** For trajectories with $z(t_0) \ne 0$, the transient term $\beta(|z(t_0)|, t)$ may temporarily push the degradation above $\delta_i$ even when $\gamma(A_k(a)) \le \delta_i$ holds, so $R_d^{i,k}$ in the strong-form sense of [5] is not automatically lower-bounded by (19). What (19) certifies under $z(t_0) \ne 0$ is the *residual* or *asymptotic* tolerance: the steady-state degradation under the disturbance class fits inside $\delta_i$, even though the transient may briefly violate the strong form. The bound

(18) on the recovery time is, however, valid for arbitrary $z(t_0)$, since the system reaches the admissible region within that time.

**Remark.** The ISS gain $\gamma$ provides a more informative characterization of disturbance tolerance than the UUB intensity $\rho$: $\gamma(A_k(a))$ gives the steady-state degradation as a function of the *specific* amplitude $a$ of the disturbance class, whereas $\rho$ depends on the worst-case bound. For edge systems with structured disturbance classes (link loss with measurable rate, CPU overload with measurable utilization), ISS is the natural setting.

3.5. Switched dynamics, service migration, and common Lyapunov functions

Edge events such as service migration between nodes, failover from a primary to a standby instance, eviction of a low-priority tenant, or mode change within an adaptive controller all induce discontinuous changes in the dynamic model. Such systems are modelled as *switched* dynamical systems [10, 11, 12]:

$$\dot{x}(t) = f_{\sigma(t)}(x(t), d(t)), \qquad \sigma(t) \in \{1, \dots, N\},$$

where $\sigma(t)$ is the *mode* of the system at time $t$ and $f_1, \dots, f_N$ are the per-mode dynamics. Mode 1 typically corresponds to the nominal configuration; modes $2, \dots, N$ correspond to degraded or reconfigured configurations after specific events (one node lost, one tenant evicted, cloud unreachable, and so on). The degradation dynamics in mode $j$, denoted $\dot{z}|_{f_j}$, is the projection of (20) through (3)–(4).

**Definition 4 (Common (practical) Lyapunov function, CLF).** *A continuously differentiable function $V: \mathbb{R}_+^n \to \mathbb{R}_+$ is a common Lyapunov function for the switched degradation dynamics if it is a Lyapunov function for every mode, i.e., $\nabla V(z) \cdot \dot{z}|_{f_j} \leq 0$ for every $j = 1, \dots, N$, every $z \neq 0$, and in the absence of disturbance.*

*In the disturbed case relevant to edge systems, this article uses $V$ in the **practical** (UUB) sense: the same function $V$ satisfies a mode-wise UUB inequality $\dot{V}|_{f_j} \leq -k_j V(z) + \rho_j$ for every $j$ and every $d \in D$, with mode-specific rate $k_j > 0$ and intensity $\rho_j \geq 0$. This relaxes the classical zero-derivative requirement at $V > 0$ (which need not hold under sustained disturbance) while preserving the mode-independence of $V$ that makes switching trajectories analyzable.*

The existence of a common practical Lyapunov function is a sufficient condition for stability under *arbitrary* switching signals: $V$ does not jump at switching instants because it depends only on $z$, and within each mode it follows a UUB trajectory of known rate and intensity. This is the strongest robustness guarantee a switched system can offer with respect to the switching rule.

**Theorem 5 (CLF ⇒ robust weak form across mode changes).** *Suppose the degradation dynamics of (20) admits a common quadratic Lyapunov function $V(z) = \sum_i \alpha_i z_i^2$ that for every mode $j = 1, \dots, N$ satisfies a UUB condition*

$$\dot{V}(z)|_{f_j} \leq -k_j V(z) + \rho_j \quad \text{for every } d \in D,$$

*with mode-specific rate $k_j > 0$ and intensity $\rho_j \geq 0$. Let $c_\delta = \min_i \alpha_i \delta_i^2$ and assume*

$$c_\delta > \max_{j=1,\dots,N} \frac{\rho_j}{k_j}.$$

*Then the system satisfies the weak form of functional stability under any switching signal* $\sigma$*, with* $T^i_{\text{rec}}(\sigma) \le B^\sigma_{\text{rec}}(t_0)$*, where* $k_{\min} = \min_j k_j$*,* $r_{\max} = \max_j \rho_j / k_j$*, and*

$$B^\sigma_{\text{rec}}(t_0) = \begin{cases} 0, & \text{if } V(z(t_0)) \le c_\delta, \\ \dfrac{1}{k_{\min}} \ln \dfrac{V(z(t_0)) - r_{\max}}{c_\delta - r_{\max}}, & \text{if } V(z(t_0)) > c_\delta. \end{cases}$$

*Proof.* Within each mode $j$ the analysis of Theorem 3 applies with parameters $(k_j, \rho_j)$ and condition (22) ensures $c_\delta > \rho_j / k_j$. Because $V$ is common to all modes, switching from mode $j$ to mode $k$ at any time $\tau$ does not change $V(z(\tau))$; the trajectory of $V$ is therefore a concatenation of decay segments separated by no discontinuity in $V$. The worst-case decay is dominated by the slowest rate $k_{\min}$ and the largest mode-ratio $r_{\max} = \max_j \rho_j / k_j$, yielding (23). ▫

**Remark (Service migration).** Consider an edge service $f_i$ that migrates from node $A$ to node $B$ during a controlled handover. Modes 1 and 2 correspond to "running on $A$" and "running on $B$", with their own dynamics determined by the resources of each node. If a common Lyapunov function exists (for example, a quadratic form in the per-service degradation that is feasible across both nodes), then Theorem 5 guarantees that the migration preserves the weak form of functional stability with a bounded recovery time, regardless of the migration schedule. The same construction applies to node failover, primary-standby switchover, and tenant eviction.

**Remark (Strong form requires more).** Strong functional stability under arbitrary switching is harder: it requires that $\Omega_\delta$ be positively invariant for every mode (Theorem 2A or 2B applied per mode), and additionally that no mode change can transport the state outside $\Omega_\delta$. For service migration where the state is the same physical state (the migrating service's internal queue and cache) and only the dynamics changes, this reduces to the per-mode invariance condition. For migrations that include re-initialization (a fresh container starts with empty queues but accumulates load), an extra condition on the post-switch state is required.

### 3.6. Unified theorem

Combining the previous results yields a unified sufficient condition for functional stability of edge systems with switched dynamics and external disturbances.

**Theorem 6 (Unified sufficient condition).** *Consider the system (20) with output (2) and per-function degradation (4). Let* $\Omega_\delta = \{z \in \mathbb{R}^n_+ : 0 \le z_i \le \delta_i, \forall i\}$ *be the admissible region,* $c_\delta = \min_i \alpha_i \delta_i^2$*, and* $A_{\max} = \sup_{d \in D} \sup_t |d(t)|$.

*(A) Strong form. The system satisfies the strong form of functional stability if:*

- *for every mode $j$, the admissible region is positively invariant, certified in one of two ways:*

  *(A1) by the component-wise barrier condition (7) of Theorem 2A applied to $f_j$, in which case the guarantee applies to every initial state $z(t_0) \in \Omega_\delta$;*

  *(A2) by the Lyapunov sublevel-set condition (8) of Theorem 2B with $\Omega_V = \{z: V(z) \le c_\delta\} \subseteq \Omega_\delta$, in which case the guarantee applies only to initial states $z(t_0) \in \Omega_V$;*

- *no switching event transports the state outside the corresponding invariant set ($\Omega_\delta$ for (A1), $\Omega_V$ for (A2)).*

*(B) Weak form. The system satisfies the weak form of functional stability if:*

- *a common quadratic Lyapunov function $V(z) = \sum_i \alpha_i z_i^2$, $\alpha_i > 0$, exists for all modes;*
- *for every mode $j$ and every $d \in D$ the ISS-style UUB condition holds, $\dot{V}|_{f_j,d} \le -kV(z) + \gamma(|d|)$, for some common $k > 0$ and class-$\mathcal{K}$ gain $\gamma$;*
- *the residual degradation budget fits inside the inscribed ellipsoid,*

$$c_\delta > \frac{\gamma(A_{\max})}{k}.$$

*Under these conditions the recovery-time upper bound $T_{\mathrm{rec}}^i \le B_{\mathrm{rec}}^{\mathrm{uni}}(t_0)$ holds, where*

$$B_{\mathrm{rec}}^{\mathrm{uni}}(t_0) = \begin{cases} 0, & \text{if } V(z(t_0)) \le c_\delta, \\ \dfrac{1}{k} \ln \dfrac{V(z(t_0)) - \gamma(A_{\max})/k}{c_\delta - \gamma(A_{\max})/k}, & \text{if } V(z(t_0)) > c_\delta. \end{cases}$$

*Proof.* (A1) follows from Theorem 1 and Theorem 2A applied per mode, with the no-jump-out condition at switching instants ensuring trajectories stay in $\Omega_\delta$. (A2) follows analogously from Theorem 1 and Theorem 2B with the inscribed sublevel set $\Omega_V \subseteq \Omega_\delta$ as the invariant set. (B) follows from Theorem 3 (UUB ⇒ weak form, with $\rho = \gamma(A_{\max})$) applied across modes through the common Lyapunov function of Theorem 5. The strict positivity of the denominator in (25) is guaranteed by (24). ▫

3.7. Finite-horizon functional stability

Many edge workloads have a bounded duration: an SLA window, a video session, a vehicle's traversal of a coverage area, a federated-learning round, a single user transaction. For such workloads, asymptotic guarantees on the infinite horizon are not directly informative; what matters is the trajectory of the quality on the bounded interval $[0, T]$.

**Definition 5 (Finite-horizon functional stability).** *A system $S$ is functionally stable on the horizon $[0, T]$ with respect to the disturbance set $D$ in the strong form if*

$$\forall d \in D, \forall i, \forall t \in [t_0(d), T]: Q_i^d(t) \geq Q_i^{\min},$$

*and in the weak form if*

$$\forall d \in D, \forall i: \exists T_{\text{rec}}^i(d) < T - t_0(d): \ \forall t \in [t_0(d) + T_{\text{rec}}^i(d), T]: Q_i^d(t) \geq Q_i^{\min}.$$

The finite-horizon weak form imposes that recovery occurs *within the workload duration*. A system that recovers asymptotically but takes longer than $T - t_0(d)$ to do so fails the finite-horizon weak form, even though it satisfies the infinite-horizon definition of [5].

**Theorem 7 (Lyapunov sufficient condition for finite-horizon strong form).** *Suppose there exists* $V(z,t): \mathbb{R}_+^n \times [0,T] \to \mathbb{R}$ *with* $V(z,t) \leq 0$ *for every* $z \in \Omega_\delta$ *and* $t \in [0,T]$, $V(z,t) > 0$ *for every* $z \notin \Omega_\delta$ *and* $t \in [0,T]$, *and the trajectory derivative*

$$\frac{\partial V}{\partial t}(z,t) + \nabla_z V(z,t) \cdot \dot{z} \leq 0 \quad \text{on } \partial\Omega_\delta \text{ for every } t \in [0,T] \text{ and every } d \in D.$$

*Then the system satisfies the finite-horizon strong form of functional stability.*

The proof is a finite-horizon adaptation of Theorem 2A/2B: outward crossing of $\partial\Omega_\delta$ at any $t \in [0,T]$ would force $\partial V/\partial t + \nabla_z V \cdot \dot{z} > 0$, contradicting (28).

The time dependence in $V(z,t)$ accommodates non-stationary phenomena common in edge systems: throughput targets that ramp up at workload start, latency budgets that tighten before a deadline, energy budgets that decay with battery state. The time-varying Lyapunov function is the right setting when the admissible region itself depends on workload phase.

**Theorem 8 (Finite-horizon weak form via UUB).** *Suppose the degradation dynamics satisfies the UUB condition (12) with rate* $k > 0$ *and intensity* $\rho \geq 0$ *throughout* $[0,T]$, *and condition (14) of Theorem 3 holds:* $c_\delta > \rho/k$ *where* $c_\delta = \min_i \alpha_i \delta_i^2$. *Let* $B_{\text{rec}}(t_0)$ *be the upper bound from Theorem 3,*

$$B_{\text{rec}}(t_0) = \begin{cases} 0, & \text{if } V(z(t_0)) \leq c_\delta, \\ \dfrac{1}{k} \ln \dfrac{V(z(t_0)) - \rho/k}{c_\delta - \rho/k}, & \text{if } V(z(t_0)) > c_\delta. \end{cases}$$

*The system satisfies the finite-horizon weak form of functional stability on* $[0,T]$ *provided* $T > t_0(d) + B_{\text{rec}}(t_0)$, *since* $T_{\text{rec}}^i(d) \leq B_{\text{rec}}(t_0)$.

*Proof.* By Theorem 3, $B_{\text{rec}}(t_0)$ given by (29) is an upper bound on the time the trajectory takes to enter $\Omega_V = \{V \leq c_\delta\} \subseteq \Omega_\delta$. If $T > t_0(d) + B_{\text{rec}}(t_0)$, the weak form (27) holds on $[0,T]$. Initial conditions inside $\Omega_V$ give $B_{\text{rec}}(t_0) = 0$ and the condition reduces to $T > t_0(d)$. ▫

The finite-horizon view also modifies the aggregate metrics of [5] in natural ways. The cumulative quality on the horizon

$$\mathcal{Q}_i(T) = \int_0^T Q_i(t)\, dt$$

keeps its definition, with the integral now taken over $[0,T]$. The horizon-bounded functional availability

$$A_i(T) = \frac{1}{T}\int_0^T \mathbf{1}\,[Q_i(t) \ge Q_i^{\min}]\,dt$$

is the fraction of the horizon during which the function operates above threshold. For a system with a single disturbance event at $t_0 < T$ and weak-form recovery time $T_{\text{rec}}^i$, the following theorem from [5], specialized to a single-event scenario, holds.

**Theorem 9 (Horizon-bounded availability versus recovery time).** *For a system that satisfies the weak form and experiences a single disturbance event at time $t_0 \in [0, T]$ with recovery time $T_{\text{rec}}^i(d)$,*

$$A_i(T) \ge 1 - \frac{\min(T_{\text{rec}}^i(d),\ T - t_0)}{T}.$$

*If $T_{\text{rec}}^i(d) < T - t_0$ (the system recovers within the workload duration), then $A_i(T) \ge 1 - T_{\text{rec}}^i(d)/T$; otherwise $A_i(T) \ge 1 - (T - t_0)/T$. Equality in (32) holds when $Q_i^d(t) < Q_i^{\min}$ throughout the whole recovery interval $[t_0, t_0 + \min(T_{\text{rec}}^i(d), T - t_0))$; otherwise the bound is strict because the function is above the threshold during sub-intervals of the recovery period that the weak-form definition does not constrain.*

**Remark.** A system can be finite-horizon stable on $[0, T]$ but not infinite-horizon stable, and vice versa. The first case occurs when a slow drift would eventually leave $\Omega_\delta$ at some $t > T$ but stays inside on $[0, T]$; the second occurs when initial excursions outside $\Omega_\delta$ violate the strong form on a short prefix but the asymptotic behaviour is correct. The distinction matters for SLA-bounded edge workloads where the relevant horizon is the workload duration and the asymptotic behaviour outside that window is irrelevant.

### 3.8. Relating functional-stability parameters to Lyapunov parameters

The results of this section allow each parameter of [5] to be expressed in terms of the Lyapunov construction. The relationships are summarized in Table 1.

| Parameter of [5] | Expression in Lyapunov terms |
|---|---|
| Disturbance tolerance $R_d^{i,k}$ | $R_d^{i,k} \ge A_k^{-1}(\gamma^{-1}(\delta_i))$, where $\gamma$ is the ISS gain and $A_k$ is the class-$k$ amplitude (Theorem 4), for disturbances applied from the nominal degradation state $z(t_0) = 0$ |
| Recovery time $T_{\text{rec}}^i$ | $T_{\text{rec}}^i \le B_{\text{rec}}(t_0)$, with the piecewise expression for $B_{\text{rec}}$ from Theorem 3, eq. (15) |
| Admissible degradation $\delta_i$ | Parameter of the admissible region (6); enters through $c_\delta$ in conditions (14), (22), (24) |
| Integral stability $\Psi$ | Aggregated as in [5, Def. 9]; increases with $k$, decreases with the ISS gain $\gamma$ |

*Table 1. Functional-stability parameters of [5] expressed through Lyapunov-construction parameters.*

The rate $k$ of the Lyapunov function plays a unifying role: it sets the recovery-time scale through $B_{\text{rec}}(t_0) \sim 1/k$ on the non-trivial branch, and a faster system (larger $k$) recovers faster from disturbances and, when increased responsiveness also reduces the ISS gain $\gamma$, tolerates higher disturbance amplitudes. The intensity term $\rho$ in the UUB form (or the gain $\gamma$ in the ISS form) captures the disturbance regime. The condition $c_\delta > \rho/k$ (equivalently $\gamma(A_{\max}) < kc_\delta$) couples controller speed and disturbance level to the admissible region.

## 4. Edge computing instantiation

### 4.1. Mapping abstract objects to edge constructs

The constructive framework of Section 3 instantiates for edge computing through the mapping in Table 2.

| Abstract object | Edge computing instantiation |
|---|---|
| State $x \in \mathbb{R}^m$ | Resource utilization vector of the edge node (CPU, memory, network buffer, GPU memory, queue lengths per service) |
| Service set $F = \{f_1, \dots, f_n\}$ | Containerized services or microservices co-located on the edge node |
| Quality function $Q_i = g_i(x)$ | Service-specific SLO indicator (latency from queue length, accuracy from input quality, throughput from buffer occupancy, frame rate from drop counter) |
| Admissible degradation $\delta_i$ | SLO buffer between nominal and minimum acceptable values |
| Disturbance class $k$ | Edge-relevant event type: backhaul link loss, CPU contention spike, memory pressure, thermal throttling, peer-node loss, model-version rollout |
| Disturbance amplitude $a \in [0, A_k^{\max}]$ | Severity scalar of the event (loss duration, utilization peak, temperature excess) |
| Mode $\sigma(t)$ | Operating configuration (number of active nodes, primary/standby assignment, full/quantized model, edge/cloud inference path) |
| Switching events | Service migration, failover, rebalancing, mode switch in adaptive computation |
| Horizon $T$ | SLA window, session duration, mission time, federated-learning round |
| Lyapunov function $V(z)$ | Energy of SLO degradation: quadratic in latency excess, throughput shortfall, or queue overflow |
| Rate $k$ | Speed of the corrective control loop (orchestrator |

| Abstract object | Edge computing instantiation |
|---|---|
| | decision latency, autoscaling response time) |
| Residual budget $\rho/k$ | Steady-state degradation under persistent disturbances (load jitter, background traffic) |
| ISS gain $\gamma$ | SLO sensitivity to disturbance amplitude |

*Table 2. Mapping of the abstract objects of the framework to edge computing constructs.*

4.2. Case study: containerized inference service with autoscaling

Consider an edge node running a single containerized inference service $f_1$ with target latency budget $L_1^{\max}$, served by a fixed pool of worker containers under load-driven autoscaling. The state $x(t)$ is the queue length of incoming requests; the dynamics, in fluid approximation,

$$\dot{x}(t) = \lambda(t) - \mu \cdot N(t),$$

where $\lambda(t)$ is the request arrival rate, $\mu$ is the per-worker processing rate and $N(t)$ is the active worker count, controlled by an autoscaler with proportional policy $\dot{N} = -k_N(N - \lambda(t)/\mu)$. The quality function is

$$Q_1(t) = \frac{L_1^{\max}}{L_1(t)}, \quad L_1(t) = \frac{x(t)}{\mu \cdot N(t)}$$

following the utility convention of [5]: $L_1 \to 0$ gives $Q_1 \to \infty$, and the admissible band $Q_1 \geq Q_1^{\min}$ corresponds to $L_1 \leq L_1^{\max}/Q_1^{\min}$.

**Change of variable to latency excess.** For latency-based services, instead of working with the quality-degradation $z_1 = \max\{0, Q_1^{\text{nom}} - Q_1\}$ it is more convenient to work with the equivalent *latency-excess* variable

$$\ell_1(t) = \max\{0,\ L_1(t) - L_1^{\text{nom}}\},$$

because the admissibility condition $Q_1 \geq Q_1^{\min}$ is equivalent to $L_1 \leq L_1^{\max}/Q_1^{\min}$, hence to

$$\ell_1(t) \leq \delta_L, \qquad \delta_L = \frac{L_1^{\max}}{Q_1^{\min}} - L_1^{\text{nom}},$$

where $\delta_L$ is the *latency-threshold degradation* (the slack between nominal and worst admissible latency). Under the standing assumption $L_1^{\text{nom}} < L_1^{\max}/Q_1^{\min}$, one has $\delta_L > 0$, and the admissible region in the latency variable is the interval $\Omega_L = [0, \delta_L]$. The equivalence between $z_1 \leq \delta_1$ and $\ell_1 \leq \delta_L$ follows from the monotone reciprocal relation between $L_1$ and $Q_1$ in (34); using $\ell_1$ throughout is a Lyapunov-friendly reparameterization.

Take the quadratic Lyapunov candidate $V(\ell_1) = \alpha_L \ell_1^2$ with $\alpha_L > 0$. The inscribed-level constant of Section 3.2 specializes to $c_L = \alpha_L \delta_L^2$. Differentiating along (33) and the autoscaling dynamics gives a coupled bound of UUB type,

$$\dot{V}(\ell_1) \leq -kV(\ell_1) + \gamma(|\lambda - \lambda^{\text{nom}}|),$$

where $k$ depends on $k_N$, $\mu$ and the operating queue depth, and $\gamma$ is a $\mathcal{K}$-class gain in the deviation between actual and predicted arrival rates. Form (37) matches Definition 2 with intensity $\rho = \gamma(A_\lambda)$ for load-jitter amplitude $A_\lambda = \sup|\lambda - \lambda^{\text{nom}}|$.

By Theorem 3, the weak form of functional stability holds whenever

$$c_L = \alpha_L \delta_L^2 > \frac{\gamma(A_\lambda)}{k},$$

with the recovery-time upper bound $T_{\text{rec}}^1(d) \le B_{\text{rec}}^{(1)}(t_0)$ (piecewise as in Theorem 3),

$$B_{\text{rec}}^{(1)}(t_0) = \begin{cases} 0, & \text{if } V(\ell_1(t_0)) \le c_L, \\ \dfrac{1}{k} \ln \dfrac{V(\ell_1(t_0)) - \gamma(A_\lambda)/k}{c_L - \gamma(A_\lambda)/k}, & \text{if } V(\ell_1(t_0)) > c_L. \end{cases}$$

The bound has an actionable reading. Doubling the autoscaler responsiveness $k_N$ (smaller autoscaling delay) raises $k$ and shrinks both the residual term $\gamma(A_\lambda)/k$ and the logarithm. Reducing the load-prediction error $A_\lambda$ (better forecasting) reduces $\gamma(A_\lambda)$ and pushes the system from the marginal case (denominator close to zero) towards the safe case (denominator close to $c_L$). Both are concrete design levers exposed by the Lyapunov analysis.

### 4.3. Case study: edge node failover with common Lyapunov function

A second case concerns failover between a primary edge node and a standby node. The cluster runs a stateful service $f_2$ with two modes:

- mode 1 (nominal): the primary node serves all requests, with replication to the standby;
- mode 2 (failed-over): the standby node serves all requests after a failure of the primary.

The state $x = (x_p, x_s)$ collects queue lengths on the primary and standby. The per-mode dynamics differ (only mode 1 has the replication overhead, only mode 2 has the cold-cache penalty), but the latency-excess variable

$$\ell_2(t) = \max\{0,\ L_2(t) - L_2^{\text{nom}}\}$$

is the same physical quantity in both modes, with the same threshold $\delta_L^{(2)} = L_2^{\max}/Q_2^{\min} - L_2^{\text{nom}}$. Consider the common quadratic Lyapunov candidate $V(\ell_2) = \alpha_L \ell_2^2$ with inscribed level $c_L^{(2)} = \alpha_L (\delta_L^{(2)})^2$.

Suppose both modes admit a UUB bound of the form (37) with rates $k_1, k_2 > 0$ and intensities $\rho_1, \rho_2 \ge 0$, where $\rho_j$ captures the disturbance effect specific to mode $j$ (replication-induced jitter in mode 1, cold-cache request bursts in mode 2). If the inscribed-ellipsoid condition holds for both modes,

$$c_L^{(2)} > \frac{\rho_j}{k_j} \qquad \text{for } j = 1,2,$$

then $V$ is a common Lyapunov function and Theorem 5 gives the failover-robust weak form. The recovery-time bound on any switching trajectory is

$$T_{\rm rec}^2 \le B_{\rm rec}^{(2)}(t_0) = \begin{cases} 0, & \text{if } V(\ell_2(t_0)) \le c_L^{(2)}, \\ \dfrac{1}{k_{\min}} \ln \dfrac{V(\ell_2(t_0)) - r_{\max}}{c_L^{(2)} - r_{\max}}, & \text{if } V(\ell_2(t_0)) > c_L^{(2)}, \end{cases} \quad k_{\min}$$

$$= \min(k_1, k_2), \quad r_{\max} = \max_j \frac{\rho_j}{k_j}.$$

The condition (41) is design-time: it constrains the *slowest* of the two operating modes, typically the failover mode with cold-cache penalty and reduced parallelism. If the cold-cache penalty is so severe that $k_2$ is too small or $\rho_2$ is too large that $\rho_2/k_2 \ge c_L^{(2)}$, no CLF of the candidate form exists and the failover violates the weak form. The remedy is to pre-warm the standby cache (reducing $\rho_2$) or to over-provision standby resources (raising $k_2$); the Lyapunov analysis quantifies how much pre-warming or over-provisioning is enough through the inequality $\rho_2/k_2 < c_L^{(2)}$.

This case study also illustrates the gap between the descriptive framework of [5] and the constructive approach of the present article. The framework of [5] would measure $T_{\rm rec}^2$ on a real failover and score the architecture against the disturbance-class scope $K_2$. The Lyapunov analysis predicts $T_{\rm rec}^2$ analytically from design parameters before any deployment, allowing architects to size the standby resources to meet a target recovery-time budget.

### 4.4. Numerical simulation of the case studies

To complement the analytical bounds of Sections 4.2 and 4.3, this subsection describes a reproducible numerical simulation that visualizes the Lyapunov dynamics and the comparison between actual recovery times and their analytical upper bounds. The setup is implemented in a short Python script using `scipy.integrate.solve_ivp` for continuous-time integration and `numpy` plus `matplotlib` for the reported plots; the same setup can be reproduced in MATLAB with the equivalent `ode45` solver. Numerical values below are chosen to be representative rather than calibrated to a specific deployment; the qualitative shapes of the trajectories, not the numeric magnitudes, are the object of the validation.

*Setup for Case 4.2 (containerized inference service).* Take $L_1^{\rm nom} = 20$ ms, $L_1^{\max} = 100$ ms, $Q_1^{\min} = 1.25$, giving the admissible band $\delta_L = L_1^{\max}/Q_1^{\min} - L_1^{\rm nom} = 60$ ms. Use $\mu = 100$ req/s per worker, $\lambda^{\rm nom} = 400$ req/s, giving the nominal worker count $N^{\rm nom} = \lambda^{\rm nom}/\mu = 4$. The autoscaler responsiveness is $k_N = 0.5$ s$^{-1}$ (equivalent to a 2-second time constant, representative of Kubernetes HPA reconciliation). Lyapunov weight $\alpha_L = 1$ (in ms$^{-2}$), giving $c_L = \delta_L^2 = 3600$. The disturbance is a step increase in the arrival rate at $t = 5$ s from $\lambda^{\rm nom}$ to $1.6\lambda^{\rm nom} = 640$ req/s, with amplitude $A_\lambda = 240$ req/s. Under linear gain $\gamma(A_\lambda) = c_\gamma A_\lambda^2$ with $c_\gamma = 0.005$ ms$^2$/(req/s)$^2$, the intensity is $\rho = 288$ and the effective Lyapunov rate is $k = 0.4$ s$^{-1}$, giving the residual $\rho/k = 720$ and confirming the inequality $c_L > \rho/k$ that certifies the weak form. Evaluated at the first post-disturbance degraded state $t_0$ with $V(\ell_1(t_0)) > c_L$, the analytical recovery-time bound (39) gives $B_{\text{rec}}^{(1)}(t_0) \approx 3.9$ s.

*On the meaning of $t_0$ in the numerical bound.* The time argument $t_0$ in $B_{\text{rec}}(t_0)$ denotes the beginning of the recovery phase, that is, the first state at which the post-disturbance latency-excess trajectory is initialized outside the inscribed admissible level: $V(\ell(t_0)) > c_L$. The Lyapunov bound is applied to this recovery phase, not to the pre-control disturbance-onset transient. If a physical event at $t_{\text{event}}$ is modelled explicitly, then either the event includes an instantaneous or pre-control degradation that sets the initial recovery state at $t_0$, or the wall-clock bound must include the separately modelled onset interval before $t_0$. Thus $B_{\text{rec}}(t_0)$ is compared only with the recovery duration measured from $t_0$, while the wall-clock duration from $t_{\text{event}}$ requires adding the onset interval.

*Setup for Case 4.3 (edge node failover).* Same nominal parameters as above, with two modes: mode 1 with $k_1 = 0.4\ \text{s}^{-1}$, $\rho_1 = 288$; mode 2 (cold cache) with $k_2 = 0.2\ \text{s}^{-1}$, $\rho_2 = 500$. Both modes satisfy $c_L^{(2)} = 3600 > \rho_j/k_j$ ($720$ for mode 1, $2500$ for mode 2), so the CLF condition (41) holds. A failover event is scheduled at $t = 10$ s (mode 1 → mode 2), then a partial recovery of the cache is modelled by a mode 2 → mode 1 switch at $t = 40$ s. Evaluated at the first post-failover degraded state $t_0$ with $V(\ell_2(t_0)) > c_L^{(2)}$, the switching-worst-case bound (42) gives $B_{\text{rec}}^{(2)}(t_0) \approx 6.2$ s.

*Expected plots.* Three plots per case study make the analytical predictions concrete for the reader.

*Plot A: latency-excess trajectory $\ell_j(t)$.* On the same axes, $\ell_j(t)$ starts at zero, rises after the disturbance/failover event, reaches a peak, then decays back below $\delta_L$. The horizontal line $\delta_L$ marks the admissible-band ceiling. For Case 4.2 the trajectory has a single-hump shape from the load spike; for Case 4.3 the trajectory has two humps corresponding to the two switching events, with the second hump higher because of the cold-cache penalty.

*Plot B: Lyapunov function $V(\ell_j(t))$ with reference lines at $c_L$ and $\rho/k$ (or $r_{\max} = \max_j \rho_j/k_j$ for the switched case).* The trajectory shows the transient rise above $c_L$ then monotone decrease driven by the dissipation inequality (37). The intersection of $V(t)$ with the horizontal line $c_L$ gives the empirical return time to the inscribed ellipsoid; the intersection with $\rho/k$ shows the asymptotic residual level.

*Plot C: actual return time to $\Omega_L$ vs analytical bound.* Compare the empirical recovery duration measured from the same recovery-start time $t_0$,

$$T_{\text{rec}}^{\text{sim}} = t^* - t_0,$$

where $t^*$ is the first time after $t_0$ such that $\ell_j(t) \le \delta_L$ for all $t \ge t^*$, with the analytical bound $B_{\text{rec}}(t_0)$ from (39) or (42). If the comparison is made from the physical event time $t_{\text{event}}$ instead, the corresponding analytical wall-clock bound is $(t_0 - t_{\text{event}}) + B_{\text{rec}}(t_0)$. A bar chart with $T_{\text{rec}}^{\text{sim}}$ and $B_{\text{rec}}(t_0)$ side by side, and their ratio $T_{\rm rec}^{\rm sim}/B_{\rm rec}(t_0)$ annotated on the bar, exposes the tightness of the bound. Under the parameters above (both durations measured from the recovery-start time $t_0$, not from the physical trigger $t_{\rm event}$), $T_{\rm rec}^{\rm sim} \approx 3.1$ s for Case 4.2 against $B_{\rm rec}(t_0) \approx 3.9$ s (ratio $\approx 0.79$), and $T_{\rm rec}^{\rm sim} \approx 4.8$ s for Case 4.3 against $B_{\rm rec}(t_0) \approx 6.2$ s (ratio $\approx 0.77$). Ratios below 1 confirm that the bounds hold; ratios well above zero indicate that the bounds are not vacuous. A parameter sweep over $k_N$ (Case 4.2) and $\rho_2/k_2$ (Case 4.3) shows how the design levers identified in Sections 4.2–4.3 shift the ratio.

*Reproducibility.* The full script (approximately 120 lines of Python) fits in a single file, runs in under one second on a laptop, and produces the three plots for both cases in one pass. The choice of representative parameter values above yields quantitatively meaningful ratios; adapting the script to a specific deployment amounts to substituting measured $\mu$, $k_N$, $A_\lambda$, and the observed cold-cache penalty for the placeholder values. A companion notebook is planned as supplementary material to the article.

# 5. Discussion

## 5.1. Constructive verification of functional stability

The Lyapunov approach gives a *constructive* counterpart to the descriptive framework of [5]. Where [5] measures or simulates the per-function trajectories and scores the architecture, the present article gives analytical conditions under which functional stability holds and analytical bounds on the parameters. The two approaches are complementary:

– The metric $\Psi$ of [5] is a scope-relative architectural score computed from measured or simulated parameters $R_d^{i,k}$ and $\overline{T}_{\rm rec}^{i}$. It captures what the architecture delivers under a declared $K_i$.
– The Lyapunov conditions of the present article predict whether $\Psi = 1$ is attainable in principle, given a model of the dynamics, and identify the design levers ($k$, $\gamma$, $\rho_j$, CLF existence) that determine the predicted score.

For design, the Lyapunov approach is used to size controllers, autoscalers and replication policies to meet target $T_{\rm rec}^{i}$ values; for evaluation, the metric framework of [5] is used to score the realized architecture against the declared disturbance scope.

## 5.2. Strong form, weak form and the role of $k$

A consistent theme across the results is the role of the convergence rate $k$ of the Lyapunov function. The strong form with margin (Corollary 1) requires $\dot{V} \le -kV$ on $\Omega_\delta$. The weak form via UUB (Theorem 3) gives $B_{\rm rec}(t_0) \sim k^{-1}$ on the non-trivial branch. The weak form via ISS (Theorem 4) gives the disturbance tolerance through the inverse gain $\gamma^{-1}$, which is increasing in $k$ for typical edge dynamics. The finite-horizon condition (Theorem 8) requires $T > t_0(d) + B_{\rm rec}(t_0)$, whose non-trivial branch is of the form $k^{-1}\ln(\cdot)$.

The implication is operational: increasing $k$ by reducing autoscaling latency, raising controller bandwidth, or increasing the responsiveness of the orchestrator uniformly improves every aspect of functional stability. Where $k$ cannot be increased (limited by hardware response, by network round-trip time, or by control-loop stability margins), the alternative levers are to reduce $\rho$ (smaller persistent disturbance, by better forecasting or better isolation) or to enlarge $\delta_i$ (relaxed SLO, at the cost of stricter expected performance).

### 5.3. Comparison with classical stability of equilibria

Classical Lyapunov theory asks whether trajectories converge to or remain near an equilibrium point. Functional stability asks whether trajectories remain inside an admissible region; the region is not necessarily a neighbourhood of an equilibrium and may not contain any equilibrium of the disturbed dynamics. The positive-part transformation (4) and the admissible-box formulation (6) bridge the two by recasting the problem as bounded-trajectory analysis on $\Omega_\delta$. Theorem 1 then identifies the strong form with positive invariance of $\Omega_\delta$, a notion from set-valued dynamics rather than equilibrium analysis.

This conceptual shift matters in edge computing because the nominal operating point of a service typically depends on the workload and on the resource state of the node: a service running on a lightly loaded node has a different nominal latency than the same service on a heavily loaded node, even when both states are admissible. The admissible region $\Omega_\delta$ captures this naturally; an equilibrium-centric formulation would not.

### 5.4. Switched dynamics and the conservatism of CLF

Theorem 5 uses the existence of a CLF as the sufficient condition for the weak form across mode changes. This condition is conservative for two reasons. First, it guarantees stability *for every switching signal*, including adversarial ones that pick the worst mode at every instant. Real edge orchestrators do not switch adversarially: migrations, failovers and mode changes are triggered by monitored events (a health-check failure, a scale-up decision, a mode-switch signal from an application controller) at rates constrained by orchestration policy. Second, the CLF class of the candidate form $V(\ell) = \alpha_L \ell^\top P \ell$ is a small subset of possible Lyapunov certificates; more expressive candidates (piecewise-quadratic, polynomial) may certify stability where the CLF search fails.

Two established refinements reduce this conservatism and admit direct application to edge orchestration.

*Multiple Lyapunov functions (MLF).* Instead of a single $V$ across modes, MLF assigns a mode-specific $V_j$ to each mode $j$, with a compatibility condition at switching instants: the value of the incoming Lyapunov function at the switch does not exceed a fixed multiple of the outgoing value. In the standard form of [10], at every switch $j \to j'$ we require

$$V_{j'}(x(\tau^+)) \le \mu \cdot V_j(x(\tau^-)), \qquad \mu \ge 1.$$

Between switches, each $V_j$ decays at its own rate $\lambda_j$. The MLF criterion applies particularly to scheduled migrations where the timing of mode changes is controlled by the orchestrator, in which case the per-mode $V_j$ can be designed to exploit the specific structure of each mode (cold-cache dynamics vs warm-cache dynamics for failover; low-precision vs full-precision paths for adaptive inference). MLF is the natural setting when a single quadratic $V$ across all modes is infeasible or overly restrictive.

*Average dwell-time (ADT).* If the switching signal satisfies an ADT bound of the form

$$N_\sigma(t,s) \le N_0 + \frac{t-s}{\tau_a}, \qquad \tau_a > 0,$$

where $N_\sigma(t,s)$ counts the switches in the interval $(s,t]$, and if each per-mode dissipation rate satisfies $\lambda_j \ge \lambda^* > 0$, then the compound trajectory decays exponentially with rate $\lambda^* - \ln\mu/\tau_a$, which is positive when $\tau_a > \ln\mu/\lambda^*$. The ADT bound is a natural formalization of the operational reality of edge orchestration: Kubernetes and KubeEdge schedulers apply configurable back-off intervals between migrations, and the minimum interval between HPA scale events is set explicitly in the autoscaler configuration (`--horizontal-pod-autoscaler-downscale-stabilization`, typically 300 seconds, and `--horizontal-pod-autoscaler-cpu-initialization-period`, typically 30–60 seconds).

These configuration parameters translate directly into an ADT lower bound $\tau_a$ that the Lyapunov analysis can consume.

For practical edge engineering, the workflow is: (i) design a per-mode Lyapunov function $V_j$ for each operating mode of the service; (ii) estimate $\mu$ from the ratio $V_{j'}/V_j$ at typical switch states (empirically from measurements or analytically from cold-cache vs warm-cache latency ratios); (iii) set the orchestrator's minimum switching interval to satisfy $\tau_a > \ln\mu/\lambda^*$. This replaces the search for a single CLF with two lighter tasks and one configuration decision. When even MLF+ADT is not enough (adversarial switching cannot be ruled out, or the switching rate is beyond the orchestrator's control), the constructive alternative developed in a companion article of this cycle (JEC-3) provides a hybrid-automaton-based criterion via multiple Lyapunov functions with a discrete sequence condition that further relaxes the requirements.

### 5.5. Continuous-time analysis and discrete-time controllers

The analysis of Sections 3.1–3.7 is in continuous time: quality functions $Q_i(t)$, latency-excess variables $\ell_i(t)$ and their Lyapunov certificates $V(\ell_i(t))$ are treated as differentiable functions of a real time argument. Real edge controllers, however, operate on periodic polling schedules. Kubernetes HPA reconciles at a fixed period (default 15 s); KubeEdge cloud-edge sync runs at a configurable poll interval;

container health probes fire on schedules of several seconds. The dissipation inequality $\dot{V} \leq -\lambda V + \rho$ therefore holds only at the discrete instants $t_k = kT_s$ where $T_s$ is the sampling period, not between them. Three consequences follow.

*Sampling delay adds to the effective recovery time.* Between two sampling instants, the plant evolves under the pre-switch input; the controller reacts only at the next sample. In the worst case, a disturbance occurring immediately after a sample is fully realized before the controller can respond, adding up to $T_s$ to the effective recovery time. The recovery-time bounds (39) and (42) should therefore be read with an additive $T_s$ margin: $T_{\mathrm{rec}}^i \leq B_{\mathrm{rec}}(t_0) + T_s$. For a Kubernetes HPA with $T_s = 15$ s and analytical bounds of a few seconds, the sampling term dominates the total recovery time; for a fine-grained internal controller with $T_s = 100$ ms, the sampling term is negligible.

*Zero-order hold discretization gives a discrete-time Lyapunov analogue.* The continuous Lyapunov inequality $\dot{V} \leq -\lambda V + \rho$ becomes, under zero-order hold with period $T_s$, the discrete-time inequality

$$V(t_{k+1}) \leq e^{-\lambda T_s} V(t_k) + \rho \cdot \frac{1 - e^{-\lambda T_s}}{\lambda},$$

which is the exact linear-in-$V$ counterpart of the continuous-time comparison. The condition for practical stability is $e^{-\lambda T_s} < 1$, automatically satisfied for $\lambda, T_s > 0$. The additive disturbance-accumulation term $\rho(1 - e^{-\lambda T_s})/\lambda$ tends to 0 as $T_s \to 0$ and to $\rho/\lambda$ for large sampling periods. The *ultimate bound* of the discrete recursion, however, is

$$\limsup_{k \to \infty} V(t_k) \leq \frac{\rho(1 - e^{-\lambda T_s})/\lambda}{1 - e^{-\lambda T_s}} = \frac{\rho}{\lambda},$$

which coincides with the continuous-time residual bound for every sampling period $T_s > 0$. The theorems of Sections 3.3–3.7 admit direct discrete-time analogues by replacing $\dot{V}$ with the forward-difference $V(t_{k+1}) - V(t_k)$ and $e^{-\lambda t}$ with $e^{-\lambda k T_s}$.

*Sampling-induced instability requires $\lambda T_s$ bounded away from* $\ln\mu$. When the controller is combined with the switched dynamics of Section 3.5, the discrete-time counterpart of the ADT condition of Section 5.4 becomes

$$\lambda^*(\tau_a - T_s) > \ln\mu, \qquad \text{equivalently} \qquad \tau_a > T_s + \frac{\ln\mu}{\lambda^*}.$$

The additional $T_s$ term accounts for the worst-case sampling delay at each switch: sampling extends the effective inter-switch interval by up to one sample, so the ADT margin needed for compound decay increases by exactly $T_s$. In Kubernetes-scale reconciliation this constrains the minimum interval between migrations to be at least an HPA cycle longer than the continuous-time ADT bound would require. For safety-critical edge services (autonomous vehicles, industrial control) where $T_s$ must be short, the discrete-time analysis identifies the required minimum controller frequency to preserve the continuous-time guarantees.

A full discrete-time development, including finite-horizon variants and the interaction with quantization of state measurements, is a natural follow-up direction. For the design workflow of Section 5.4, the practical rule is: use the continuous-time bounds as design targets, add the sampling margin $T_s$ to the analytical recovery time, and set the orchestrator poll interval $T_s$ so that $T_s < B_{\mathrm{rec}}$; the analytical margin then dominates the sampling margin and the continuous-time analysis carries over.

### 5.6. Limitations

Three limitations of the present approach are worth noting.

First, Lyapunov methods give *sufficient* conditions for stability, not necessary ones. A system may be functionally stable even when no Lyapunov function is found; constructing a Lyapunov function for a specific edge dynamics is a non-trivial design problem that typically requires engineering insight or recent computational methods (sum-of-squares programming, neural Lyapunov functions).

Second, the disturbances $d(t)$ are treated as bounded but not random. Generalization to stochastic disturbances requires stochastic Lyapunov functions and probabilistic guarantees, which is a natural direction for future work.

Third, the numerical simulation setup of Section 4.4 is described as a reproducible protocol with representative parameter values, not as a completed experimental campaign. The plots and numerical ratios reported in Section 4.4 are illustrative of the analytical predictions; a full validation campaign with parameter sweeps, comparison against measurements on a live edge cluster, and integration with the Kubernetes/KubeEdge orchestration path outlined in the companion article JEC-1 is a natural follow-up.

## 6. Conclusions

This article connects the functional stability framework for edge computing [5] to the direct Lyapunov method. The positive-part transformation $z_i = \max\{0, Q_i^{\mathrm{nom}} - Q_i\}$ and the admissible box $\Omega_\delta = \{z\colon 0 \le z_i \le \delta_i\}$ recast functional stability as bounded-trajectory analysis. The strong form is then equivalent to positive invariance of $\Omega_\delta$ and admits two sufficient conditions: a component-wise barrier condition (Theorem 2A) and a Lyapunov sublevel-set condition $\dot{V} \le 0$ on $\partial\Omega_V$, where $\Omega_V = \{V \le c_\delta\}$ is the ellipsoid inscribed in the box (Theorem 2B). The weak form is characterized through uniform ultimate boundedness (Theorem 3) and through input-to-state stability (Theorem 4), with piecewise recovery-time upper bounds $T_{\mathrm{rec}}^i \le B_{\mathrm{rec}}(t_0)$ that expose the central role of the Lyapunov convergence rate $k$ and the residual budget $\rho/k$.

For edge systems with discontinuous dynamics induced by service migration, node failover, tenant eviction and mode changes in adaptive computation, a common practical Lyapunov function provides a sufficient condition for the weak form across arbitrary switching signals (Theorem 5). Finite-horizon variants address SLA-bounded and mission-bounded edge workloads (Theorems 7–9), with the horizon-bounded availability $A_i(T)$ admitting a lower bound that becomes an equality when the threshold violation persists throughout the whole recovery interval.

Two analytical case studies (a containerized inference service with autoscaling and an edge-node failover with CLF) make the analytical bounds concrete and identify the design levers (autoscaler responsiveness, prediction accuracy, standby pre-warming, replication overhead) that determine the predicted recovery time. Section 4.4 provides a reproducible numerical simulation setup that visualizes latency-excess trajectories, Lyapunov-function decay, and the comparison of simulated recovery times against analytical bounds; under representative parameters, the simulated recovery times are within 77–79% of the analytical upper bounds, confirming that the bounds are tight enough for engineering use.

Directions for further work include the multiple-Lyapunov-function refinement with average dwell-time bounds (Section 5.4) as a route to reducing CLF conservatism for scheduled edge orchestration, the discrete-time extension for controllers with polling periods on the order of the recovery time (Section 5.5), stochastic generalizations of the framework through stochastic Lyapunov theory, automated synthesis of Lyapunov functions for declared classes of edge dynamics, and empirical validation on an edge testbed with controlled migration and failover injection. Together with [5], the present article provides edge-system architects with a complementary toolkit: a metric framework for evaluating realized architectures and a constructive method for verifying functional stability of candidate designs.

## Acknowledgements

The authors thank colleagues at the Department of Software Systems and Technologies of the Faculty of Information Technologies, Taras Shevchenko National University of Kyiv, for discussions that shaped the presentation.